\documentstyle[prl,aps,preprint,eqsecnum]{revtex}
\begin{document}
\draft

\title{On the gravitational moments of a Dirac particle}
\author{Yuri N. Obukhov
\address{Department of Theoretical Physics, Moscow State University,
117234 Moscow, Russia}}
\maketitle

\abstract{We consider the classical theory of the Dirac massive particle 
in the Riemann-Cartan spacetime. We demonstrate that the translational and 
the Lorentz gravitational moments, obtained by means of the Gordon type 
decompositions of the canonical energy-momentum and spin currents, are 
consistently coupled to torsion and curvature, as expected.}

\section{Introduction}

Riemann-Cartan geometry arises naturally in the gauge theory of
the Poincar\'e group, see e.g. \cite{AT,PR}. One can interpret spacetime
coframe $\vartheta^\alpha$ and local Lorentz connection $\Gamma_\alpha
{}^\beta$ as the gravitational potentials related to the translation
group and the Lorentz group, respectively. The two-forms of torsion 
and curvature, $T^\alpha:=d\vartheta^\alpha +\Gamma_\beta{}^\alpha\wedge
\vartheta^\beta$, $R_\alpha{}^{\beta}:=d\Gamma_\alpha{}^\beta + 
\Gamma_\gamma{}^\beta\wedge \Gamma_\alpha{}^\gamma$, 
represent the corresponding gauge field strengths. 
Besides, the Riemann-Cartan spacetime carries a metric $g_{\alpha\beta}$ 
which is covariantly constant: $Dg_{\alpha\beta}$ $=0$. Usually one chooses 
$g_{\alpha\beta}=o_{\alpha\beta}$, the flat Minkowski metric, thus 
restricting oneself to orthonormal frames and coframes.

General dynamical scheme of the Poincar\'e gauge theory is well 
established. The Noether currents of matter fields, $\Sigma_\alpha$
(the canonical energy-momentum three-form) and $\tau_{\alpha\beta}$
(the canonical spin three-form), are coupled to the translational
$\vartheta^\alpha$ and the Lorentz $\Gamma^{\alpha\beta}$ gauge
potentials, respectively. These currents are thus representing the two
types of {\it gravitational charges} of a matter source. 

Recently, the Dirac electron theory has been analyzed in {\it flat Minkowski 
spacetime} \cite{HMMO}, and the structure of the canonical energy-momentum 
and spin currents was studied in detail. It was shown, 
developing analogy with electrodynamics, that a Dirac particle is naturally 
characterized by two gravitational moments. In simple physical terms,
one can describe them as the Amp\'ere type ring currents induced by
the two gravitational charges via spin of a particle. Here we consider
the Dirac theory with the gravitational field ``switched on". We demonstrate
the consistency of the coupling of the gravitational moments to the 
torsion and curvature.    

It is worthwhile to mention that gravitational moments of a Dirac particle
were discussed also by Kobzarev and Okun \cite{KO} and Khriplovich 
\cite{khrip} in the framework of Einstein's general relativity theory, 
whereas in \cite{seitz,lemke} gravitational moments of higher spins and
in lower dimensions were studied. 

In this paper, we are using the basic conventions and notations of Bjorken 
and Drell. In particular, the constant Minkowski metric is $o_{\alpha\beta} =
{\rm diag}(+1,-1,-1,-1)$, and we choose the Dirac matrices $\gamma^\alpha$ 
in the standard form of \cite{BD}. In the exterior algebra on a spacetime,
$\wedge, \rfloor$ are exterior and interior products, respectively, ${}^\ast$ 
is the Hodge star operator. Local orthonormal frame $e_\alpha$ is dual to a 
coframe $\vartheta^\alpha$ one-form: $e_\alpha\rfloor\vartheta^\beta
=\delta_\alpha^\beta$. Finally, starting from the volume one-form
$\eta:={}^\ast 1$, one defines Trautman's $\eta$-basis as usual by 
$\eta_\alpha:=e_\alpha\rfloor\eta={}^\ast\vartheta_\alpha$, 
$\eta_{\alpha\beta}:=e_\beta\rfloor\eta_\alpha ={}^\ast(\vartheta_\alpha
\wedge\vartheta_\beta)$ etc.

\section{Fermions in the Riemann-Cartan spacetime}

For the description of classical Dirac particle of mass $m$, we will
use the formalism of Clifford algebra--valued exterior forms \cite{AT}.
The following matrix-valued one- or three-forms are basic objects in this
approach:
\begin{equation}
\gamma:=\gamma_\alpha\,\vartheta^\alpha\,,\qquad{}^\ast\gamma=\gamma^\alpha
\,\eta_\alpha\,.\label{gamma}
\end{equation}
Exterior product yields a two-form:
\begin{equation}
\widehat{\sigma}:=\frac{i}{2}\,\gamma\wedge\gamma =
\frac{1}{2}\widehat{\sigma}_{\alpha\beta}\,
\vartheta^\alpha\wedge\vartheta^\beta.\label{sig}
\end{equation}
The coefficients $\widehat{\sigma}_{\alpha\beta}:= i\gamma_{[\alpha}
\gamma_{\beta]}$ generate infinitesimal Lorentz transformations of spinor
fields. Defining another constant matrix, $\gamma_5 :=-i\,\gamma^{\hat{0}}
\gamma^{\hat{1}}\gamma^{\hat{2}}\gamma^{\hat{3}}$, it is straightforward to 
prove the fundamental identities for the Clifford algebra-valued objects:
\begin{eqnarray}
\widehat{\sigma}_{\alpha\beta}\,{}^\ast\widehat{\sigma}&=& \eta_{\alpha\beta}
- i\,\gamma_5\,\vartheta_\alpha\wedge\vartheta_\beta 
- 2i\,\vartheta_{[\alpha}\wedge e_{\beta]}\rfloor{}^\ast\widehat{\sigma} 
,\label{sisi3}\\
{}^\ast\widehat{\sigma}\,\widehat{\sigma}_{\alpha\beta}&=& \eta_{\alpha\beta}
- i\,\gamma_5\,\vartheta_\alpha\wedge\vartheta_\beta 
+ 2i\,\vartheta_{[\alpha}\wedge e_{\beta]}\rfloor{}^\ast\widehat{\sigma} 
.\label{sisi4}
\end{eqnarray}

The Lagrangian four-form of a Dirac field $\Psi$ is given by 
\begin{equation}
L_{\rm D}= 
\frac{i}{2}{\hbar}\left\{\overline{\Psi}\,{}^\ast\gamma\wedge D\Psi
+\overline{D\Psi}\wedge{}^\ast\gamma\,\Psi\right\}+{}^\ast mc\,
\overline{\Psi}\Psi\,.\label{lagr}
\end{equation}
Dirac fields are local sections of the spinor $SO(1,3)$-bundle
associated with the principal bundle of orthonormal frames. Hence, the
spinor covariant derivative in the Riemann-Cartan spacetime is defined by
\begin{equation}
D\Psi:=d\Psi+\frac{i}{4}\Gamma^{\alpha\beta}\wedge 
\widehat{\sigma}_{\alpha\beta}\,\Psi,\qquad 
\overline{D\Psi}=d\overline{\Psi}-\frac{i}{4}\Gamma^{\alpha\beta}\wedge
\overline{\Psi}\widehat{\sigma}_{\alpha\beta}\,.\label{covD}
\end{equation}
The Dirac field equation, which arises from the Lagrangian (\ref{lagr}),
reads
\begin{eqnarray}
i\hbar{}^\ast\gamma\wedge \left(D\,\Psi - \hbox{$\scriptstyle\frac{1}{2}$}
T\,\Psi\right) + {}^\ast mc\,\Psi &=&0,\label{dirRCa}\\
i\hbar\left(\overline{D\Psi} - \hbox{$\scriptstyle\frac{1}{2}$}
T\,\overline{\Psi}\right)\wedge{}^\ast\gamma+
{}^\ast mc\,\overline{\Psi} &=&0\,, \label{dirRCb}
\end{eqnarray}
where $T:=e_\alpha\rfloor T^\alpha$ is the torsion trace one-form.

The standard Lagrange--Noether machinery in the gauge gravity, see e.g. 
\cite{PR}, provides a general definition of the energy-momentum and spin 
currents in non-flat spacetime (accounting also for the
possibility of non-minimal coupling and ``Pauli-type'' terms):
\begin{eqnarray}
\Sigma_\alpha &:=& e_\alpha\rfloor L-(e_\alpha\rfloor D\Psi^A)\wedge
{\frac{\partial L}{\partial D\Psi^A}}-(e_\alpha\rfloor\Psi^A)\wedge
{\frac{\partial L}{\partial\Psi^A}}\nonumber\\
&&+ D{\frac{\partial L}{\partial T^\alpha}}-
(e_{\alpha}\rfloor T^\beta)\wedge{\frac{\partial L}{\partial T^\beta}} 
- (e_{\alpha}\rfloor R_{\beta}{}^{\gamma})\wedge 
{\frac{\partial L}{\partial R_{\beta}{}^{\gamma}}},\label{momcan}\\
\tau_{\alpha\beta} 
&:=&(\ell_{\alpha\beta}{}^A_B\,\Psi^B)\wedge{\frac {\partial L}{\partial 
D\Psi^A}} + \vartheta_{[\alpha}\wedge {\frac {\partial L}{\partial 
T^{\beta]}}}+D\,{\frac {\partial L}{\partial R^{\alpha\beta}}}\,.\label{spC}
\end{eqnarray}
Here $\Psi^A$ is a set of arbitrary matter fields, with $\ell_{\alpha\beta}$
denoting the corresponding Lorentz group generators.

The first and second Noether theorems yield two covariant conservation
laws which are fullfilled on the classical matter field equations:
\begin{eqnarray}
D\Sigma_\alpha &=& 0,\\
D\tau_{\alpha\beta} + \vartheta_{[\alpha}\wedge\Sigma_{\beta]}&=&0.
\end{eqnarray}

In case of the Dirac theory under consideration, matter is described
by the pair of independent four-spinors $\Psi^A=\{\Psi,\overline{\Psi}\}$,
and the canonical gravitational currents are straightforwardly obtained
after substituting (\ref{lagr}) into (\ref{momcan})-(\ref{spC}) 
[hereafter $D_\alpha:=e_\alpha\rfloor D$]:
\begin{eqnarray}
\Sigma_\alpha &=& {i\hbar\over 2}\left(\overline{\Psi}{}^\ast\gamma
D_\alpha\Psi - D_\alpha\overline{\Psi}{}^\ast\gamma\Psi\right),\label{Sa}\\
\tau_{\alpha\beta}&=&{\hbar\over 4}\,\vartheta_\alpha\wedge\vartheta_\beta
\wedge\overline{\Psi}\gamma\gamma_5\Psi.\label{Tau}
\end{eqnarray}

\section{Gravitational moments of a Dirac particle}

In a theory invariant with respect to some {\it internal} gauge group $G$, 
the {\it generalized moment} is a $\cal G$-valued two-form such that its 
exterior differential produces the polarizational part of the related 
Noether current, with $\cal G$ denoting the Lie algebra of $G$. On the 
Lagrangian level, the moment couples directly to the gauge field strength.
To be specific, in the Dirac--Yang-Mills theory the Noether ``isospin'' 
current $J_{{}_K}=-ie\overline{\Psi}\tau_{{}_K}\Psi$ couples to the gauge 
potential one-form $A^{{}_K}$, with $e$ denoting the coupling constant, and
$\tau_{{}_K}$ being generators of the gauge group. The Gordon decomposition 
\cite{HMMO} reveals a nontrivial substructure of this coupling by relating 
the polarization moment two-form $P_{{}_K}=-i{e\hbar\over 2mc}\overline{\Psi}
\tau_{{}_K}{}^\ast\widehat{\sigma}\Psi$ directly to the gauge field strength 
$F^{{}_K}$. One may expect similar results to hold when passing from internal 
symmetry groups to spacetime symmetries. Geometrically this means a 
departure from flat Minkowski spacetime by ``switching on'' gravity.

Technically, we can proceed along the same lines as for the Dirac--Yang-Mills 
theory \cite{HMMO}. We use (\ref{dirRCa})-(\ref{dirRCb}) in order to 
express $\Psi$ and $\overline{\Psi}$ in terms of the differentials, and 
then substitute them back into the Dirac equation and into the Noether
currents (\ref{Sa})-(\ref{Tau}).  After some algebra, we find the squared 
Dirac equation in the form
\begin{equation}
(D\,{}^\ast D + 2\,S\wedge D + X)\,\Psi=0,\label{Dsq2}
\end{equation}
where the three-form $S$ and the four-form $X$ read, respectively,
\begin{eqnarray}
S&:=&-\,{\frac i 2}\,(D{}^\ast\widehat{\sigma} + 
T\wedge{}^\ast\widehat{\sigma}),\label{S}\\
X&:=&{}^\ast\!\left({mc\over\hbar}\right)^2 + {\frac 1 4}\,{}^\ast
\widehat{\sigma}\wedge R_{\alpha\beta}\,
\widehat{\sigma}^{\alpha\beta}\nonumber\\
+ {\frac 1 2}\Big(-d\,{}^\ast T\! &+&\! i\,{}^\ast\widehat{\sigma}\wedge dT - 
{\frac 1 2}\,T\wedge{}^\ast T + i\,(D{}^\ast\widehat{\sigma})\wedge T\Big).
\label{X}
\end{eqnarray}
Here we have used the Ricci identity for the spinor covariant derivative
(\ref{covD}):
\begin{equation}
DD\Psi = {\frac i 4}R_{\alpha\beta}\,\widehat{\sigma}^{\alpha\beta}\Psi.
\end{equation}
One can immediately verify that equation (\ref{Dsq2}) can be derived
from the Lagrange four-form
\begin{equation}
L_{D^2}=L^{(c)} + L^{(p)},\label{Ldec}
\end{equation}
where the {\it convective Lagrangian} and the {\it polarizational 
Lagrangian} read, respectively:
\begin{eqnarray}
L^{(c)}&:=&{1\over 2}\left({\hbar^2\over mc}\,{}^\ast\overline{D\Psi}\wedge 
D\Psi + {}^\ast mc\,\overline{\Psi}\,\Psi\right),\label{Lconv}\\
L^{(p)}&=&M_{\alpha\beta}\wedge R^{\alpha\beta} + M_\alpha\wedge T^\alpha 
- {\frac{\hbar^2}{8mc}}\,T\wedge {}^\ast T\,\overline{\Psi}\,\Psi.\label{LDp}
\end{eqnarray}
Here $M_{\alpha\beta}$ is the {\it Lorentz gravitational moment} given by 
\begin{equation}
M_{\alpha\beta}:={\hbar^2\over 16mc}\overline{\Psi}({}^\ast\widehat{\sigma}
\widehat{\sigma}_{\alpha\beta} + \widehat{\sigma}_{\alpha\beta}{}^\ast
\widehat{\sigma})\Psi, \label{lorgrav} 
\end{equation}
and
\begin{equation}
M_\alpha:= {\frac{\hbar^2}{4mc}}\Big[i\,\overline{\Psi}\,{}^\ast
\widehat{\sigma}D_\alpha\Psi - i\,D_\alpha\overline{\Psi}{}^\ast
\widehat{\sigma}\,\Psi - (e_\alpha\rfloor{}^\ast\overline{D\Psi})\,\Psi 
- \overline{\Psi}(e_\alpha\rfloor{}^\ast D\Psi)\Big].\label{Mtrans}
\end{equation}

We are now in a position to find the decosmposition of the energy-momentum 
and spin currents into the convective and polarization parts corresponding 
to the decomposition of the Lagrangian (\ref{Ldec}),
\begin{eqnarray}
\Sigma_\alpha &=& \Sigma_\alpha^{(c)} + \Sigma_\alpha^{(p)},\label{decmom}\\
\tau_{\alpha\beta} &=& \tau_{\alpha\beta}^{(c)} + 
\tau_{\alpha\beta}^{(p)}.
\end{eqnarray}
A straightforward use of the {\it convective} Lagrangian (\ref{Lconv}) 
in (\ref{momcan}) and (\ref{spC}) yields the the three-forms
\begin{eqnarray}
\Sigma_\alpha^{(c)}&:=&{mc\over 2}\,\overline{\Psi}\Psi\,\eta_\alpha
+ {\hbar^2\over 4mc}\Big[{}^\ast(\overline{D\Psi})D_\alpha\Psi
+ D_\alpha\overline{\Psi}{}^\ast D\Psi \nonumber\\
&&+ (e_\alpha\rfloor{}^\ast\overline{D\Psi})\wedge D\Psi + \overline{D\Psi}
\wedge(e_\alpha\rfloor{}^\ast D\Psi)\Big],\label{momconv}\\
\tau_{\alpha\beta}^{(c)}&=&-\,{i\hbar^2\over 8mc}\left({}^\ast
\overline{D\Psi}\widehat{\sigma}_{\alpha\beta}\Psi - \overline{\Psi}
\widehat{\sigma}_{\alpha\beta}{}^\ast D\Psi\right).\label{tauC}
\end{eqnarray}
For the {\it polarizational} part, we need the
derivatives with respect to curvature and torsion:
\begin{eqnarray}
{\frac{\partial L^{(p)}}{\partial R^{\alpha\beta}}}&=& 
M_{\alpha\beta},\label{Mab1}\\
{\frac{\partial L^{(p)}}{\partial T^\alpha}}&=& M_\alpha + {\frac {\hbar^2} 
{4mc}}\,(e_\alpha\rfloor{}^\ast T)\,\overline{\Psi}\,\Psi\,
=:{\hbox{\it \v{M}}}_\alpha.\label{Mtrans1}
\end{eqnarray}
Here we recover the correct {\it translational grativational moment}
\cite{HMMO}
\begin{equation}
{\hbox{\it \v{M}}}_\alpha = -\,{i\hbar^2\over 4mc}\,[\overline{\Psi}\,
(e_\alpha\rfloor{}^\ast\,\hat{\sigma})\wedge D\Psi + \overline{D\Psi}
\wedge(e_\alpha\rfloor{}^\ast\,\hat{\sigma})\,\Psi].\label{Mtil}
\end{equation}
In (\ref{Mtrans1}) we used the Riemann-Cartan the identity which holds 
for all spinor fields satisfying the Dirac equation:
\begin{equation}
i\left(\overline{\Psi}\,^\ast\widehat{\sigma}\wedge D\Psi - 
\overline{D\Psi}\wedge\,^\ast\widehat{\sigma}\Psi \right) = 
{}^\ast D(\overline{\Psi}\Psi) - {}^\ast T\,\overline{\Psi}\Psi.\label{MD1}
\end{equation}
Substituting (\ref{Mab1}),(\ref{Mtrans1}) into (\ref{momcan}),(\ref{spC}), 
we finally obtain the polarizational energy-momentum and spin currents 
of a Dirac particle in the Riemann-Cartan spacetime:
\begin{eqnarray}
\Sigma_\alpha^{(p)}&=&D\,{\hbox{\it \v{M}}}_\alpha+\Sigma_\alpha^{(RC)},\\ 
\tau_{\alpha\beta}^{(p)}&=&\vartheta_{[\alpha}\wedge{\hbox{\it
\v{M}}}_{\beta]} + D\,M_{\alpha\beta} + \tau_{\alpha\beta}^{(RC)}.
\end{eqnarray}
The terms $\Sigma_\alpha^{(RC)}$ and $\tau_{\alpha\beta}^{(RC)}$
contain curvature and torsion explicitly; these contributions are 
absent in flat Minkowski spacetime. For completeness, we give their
explicit form:
\begin{eqnarray}
\Sigma_\alpha^{(RC)}&=&(e_\alpha\rfloor M_{\rho\sigma})\wedge
R^{\rho\sigma} - {\hbox{\it \v{M}}}_\alpha\wedge T +
(e_\alpha\rfloor{\hbox{\it \v{M}}}_\beta - e_\beta\rfloor{\hbox{\it
\v{M}}}_\alpha)\wedge T^\beta\nonumber\\ &&-(e_\alpha\rfloor
e_\beta\rfloor U)\wedge T^\beta + (e_\alpha\rfloor U) \wedge T +
(e_\alpha\rfloor{}^\ast U)\wedge{}^\ast T\nonumber\\ &&-{\frac
{\hbar^2} {8mc}}\,\left[(e_\alpha\rfloor T)\,{}^\ast T + T\wedge
e_\alpha \rfloor{}^\ast T\right]\,\overline{\Psi}\Psi,\label{sigcur}\\ 
\tau_{\alpha\beta}^{(RC)}&=& -e_\gamma\rfloor(M_{\alpha\beta}
\wedge T^\gamma).\label{taucur}
\end{eqnarray}

As was noticed in \cite{HMMO}, the three-form 
\begin{equation}
U={1\over 2}\,\vartheta^\alpha\wedge\check M_\alpha\,=-\,{i\hbar^2\over 
4mc}\left(\overline{\Psi}\,^\ast\widehat{\sigma}\wedge D\Psi
- \overline{D\Psi}\wedge\,^\ast\widehat{\sigma}\Psi \right)\label{Udir}
\end{equation}
plays a role of a ``superpotential" from which all the moments of a
Dirac particle are generated. For comparison, it is instructive to
collect the properties of the moments in a Table~\ref{table}.
\begin{table}
\centering
\caption{Gauge couplings and moments of the Dirac particle}\label{table}
\begin{tabular}{|cc|c|c|c|c|}
\hline\hline
\multicolumn{2}{|c|}{\it Gauge model}&{\it Generator}&{\it Field}&
{\it Moment 2-form}&{\it Dim}\\      \hline\hline
\multicolumn{2}{|c|}{}&&&&\\
\multicolumn{2}{|c|}{Maxwell\ \ \ \ $U(1)$}&$i$&$F$&${e\hbar\over 2mc}
\overline{\Psi}{}^\ast\widehat{\sigma}\Psi$&$[e]$\\ 
\multicolumn{2}{|c|}{}&&&&\\     \hline
\multicolumn{2}{|c|}{}&&&&\\
\multicolumn{2}{|c|}{Yang-Mills\ \ $SU(N)$}&$\tau_{_{K}}$&$F^{_{K}}$
&$-i{e\hbar\over 2mc}\overline{\Psi}\tau_{_{K}}{}^\ast
\widehat{\sigma}\Psi$&$[e]$\\  \multicolumn{2}{|c|}{}&&&&\\ \hline 
\multicolumn{1}{|c|}{}&\multicolumn{1}{c|}{}&&&
$-i{\hbar^2\over 4mc}\{\overline{\Psi}(e_\alpha
\rfloor{}^\ast\widehat{\sigma})\wedge D\Psi$&\\ 
\multicolumn{1}{|c|}{}&\multicolumn{1}{c|}{$T_4$}&
$D_\alpha$&$T^\alpha$&&$[mc]$\\
\multicolumn{1}{|c|}{Poincar\'e}&\multicolumn{1}{c|}{}&&&
$+\overline{D\Psi}\wedge (e_\alpha\rfloor{}^\ast\widehat{\sigma})\Psi\}$&\\
\cline{2-6}
\multicolumn{1}{|c|}{gravity}&\multicolumn{1}{c|}{}&&&&\\
\multicolumn{1}{|c|}{}&\multicolumn{1}{c|}{$SO(1,3)$}&${i\over 4}
\widehat{\sigma}_{\alpha\beta}$&$R^{\alpha\beta}$
&${\hbar^2\over 16mc}\overline{\Psi}({}^\ast\widehat{\sigma}
\widehat{\sigma}_{\alpha\beta} + \widehat{\sigma}_{\alpha\beta}
{}^\ast\widehat{\sigma})\Psi$&${[\hbar]}$\\
\multicolumn{1}{|c|}{}&\multicolumn{1}{c|}{}&&&&\\
\hline\hline
\end{tabular}
\end{table}

\section{Discussion and conclusion}

In this paper, we have generalized our previous discussion \cite{HMMO}
of the gravitational moments to the case of the Riemann-Cartan spacetime.

Although the structure of the torsion- and curvature-dependent terms
(\ref{sigcur}) and (\ref{taucur}) is not physically transparent, it is
remarkable that the gravitational moments ${\hbox{\it \v{M}}}_\alpha$
and $M_{\alpha\beta}$ are the same for a Minkowski and a
Riemann-Cartan spacetime.  Moreover, the specific coupling
$\{moment\times gauge\ field\ strength\}$ on the Lagrangian level is
correctly reproduced in the Poincar\'e gauge--Dirac theory, see
(\ref{LDp}).

The results obtained are helpful for deepening our understanding of
the classical limit of the Dirac theory. A relevant general discussion
of the low-energy limit can be found in \cite{HN} (for non-inertial frames 
in flat spacetime), and in \cite{ryder} (for the Riemann-Cartan spacetime).

The present study clearly demonstrates the complete consistency of the
properties of {\it two} gravitational moments with the fundamental structure
of the Poincar\'e gauge gravity which naturally operates with the two 
types of gravitational charge, mass and spin. At the same time, the 
definition and properties of a gravitational moment in a purely Riemannian 
spacetime of Einstein's general relativity (GR) remain unclear. A somewhat 
paradoxical situation arises: only a {\it translational} gravitational
charge (mass, or energy-momentum) is available in GR, but evidently only
a {\it Lorentz} gravitational moment can survive for the case of the
vanishing torsion [recall that a moment is ``conjugated" to a corresponding
gauge field strength, cf. (\ref{Mab1})-(\ref{Mtrans1})]. This problem will
be analyzed elsewhere. 

\section{Acknowledgements}

This work was supported by Deutsche Forschungsgemeinschaft (Bonn)
under project He 528/17-2.

\end{document}